# High-Quality Stepped-Impedance Resonators Suitable for Circuit-QED Measurement of Superconducting Artificial Atoms

Yirong Jin, Hui Deng, Xueyi Guo, Yarui Zheng, Keqiang Huang, Luhui Ning, and Dongning Zheng

*Abstract*—High-quality factor coplanar resonators are critical elements in superconducting quantum circuits. We describe the design, fabrication, and measurement of stepped impedance resonators (SIRs), which are more compact in size than commonly used uniform impedance resonators (UIRs). With properly chosen impedance ratio, SIRs can be 27% shorter than UIRs. As a result, the area occupied by SIRs can be reduced. Two kinds of designs containing both SIRs and UIRs are fabricated and measured. The power dependence of the extracted internal quality factors ($Q_i$) for all the resonators showed that SIRs and UIRs have comparable performance with $Q_i$ around half-million under single-photon level excitation. These results indicate that SIRs could be used in superconducting quantum circuits. The reduced size of SIRs may also lead to reduced overall circuit area and increased integration level.

## I. Introduction

MICROWAVE resonators with high quality factor play a critical role in superconducting quantum circuits [1]–[5] and in other applications requiring high sensitivity [6]. Coplanar resonators are widely used in superconducting qubit devices for their advantages including compact size, design flexibility and simplicity, easily coupling to qubits, etc. Although the internal quality factor ($Q_i$) of coplanar resonators is usually lower than that of state-of-art 3D microwave cavities [5], coplanar resonators are still excellent choice for many purposes such as wiring up solid state quantum bits [1], [7]. By performing careful surface treatment of the substrates [8], or by deeply etching away materials between the central conductor and ground [9], $Q_i$ of superconducting coplanar resonators can reach over one million in the quantum region. In a recent report, by using coplanar

Manuscript received September 6, 2016; accepted January 23, 2017. Date of publication February 2, 2017; date of current version February 16, 2017. This work was supported in part by the National Basic Research Program of China (973 Program) under Grants 2014CB921401, 2014CB921202, and 2016YFA0300601, in part by the National Natural Science Foundation of China under Grants 91321208, 11374344, 11674376, and 11404386, and in part by the Strategic Priority Research Program of the Chinese Academy of Sciences under Grant XDB07000000. (*Corresponding author: Yirong Jin.*)

The authors are with Beijing National Laboratory for Condensed Matter Physics, Institute of Physics, Chinese Academy of Sciences, Beijing 100190, China (e-mail: jyr-king@iphy.ac.cn; hdeng@iphy.ac.cn; guoxueyi@iphy.ac.cn; zhengyarui@iphy.ac.cn; kqhuang@iphy.ac.cn; lhning@iphy.ac.cn; dzheng@iphy.ac.cn).

Color versions of one or more of the figures in this paper are available online at http://ieeexplore.ieee.org.

Digital Object Identifier 10.1109/TASC.2017.2663382

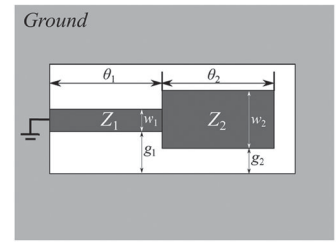

Fig. 1. A schematic of a $\lambda/4$ coplanar stepped impedance resonator with impedance ratio $R$ less than 1.

resonators, a nine qubits system has successfully demonstrated 1-D surface code quantum error correction [10], which was considered as a ground breaking step toward 2-D surface code scheme.

The most popular type of coplanar resonators used in quantum circuits is the so called uniform impedance resonator (UIR), which is consisted of a segment of coplanar transmission line opened at one or both ends. UIR requires a length of at least $\lambda/4\sqrt{\epsilon_{\text{eff}}}$. For frequencies at microwave range, this corresponds to $\sim$1 cm, which is much larger than typical dimension of a superconducting qubit [10]. Meander structure is commonly used to mitigate the area occupation. Still, resonators are usually among the largest elements in a superconducting quantum circuit. With the increased requirement for more and more entangled qubits in future development, reducing the area for a single qubit cell, including the artificial atom, the readout resonator and the control lines, become very desirable.

In this paper, we investigated another type of coplanar resonator, called stepped impedance resonator (SIR) [11]. With properly chosen parameters, SIRs can be much more compact than UIRs. We designed, fabricated and systematically measured the internal quality factors of SIRs and UIRs made either on the same chip or different chips with the same fabrication recipe. The results showed convincingly that the performance of SIRs is comparable with UIRs. Therefore, SIRs could be used to replace UIRs in various qubit devices.

SIRs have long been used in applications including microwave filter and antenna [12], [13], or RF coil in MRI [14]. The main advantages of SIRs include the compact size and improved spurious characteristics. A SIR is consisted of two parts of transmission line with different characteristic impedances, as shown in Fig. 1. The resonance condition can be derived by its zero point of input admittance [11]:

$$R = \tan\theta_1 \cdot \tan\theta_2, \quad R = \frac{Z_1}{Z_2}.$$

Here $R$ is the impedance ratio, $\theta_1$ and $\theta_2$ are the electrical lengths of the two parts of transmission line, respectively. When $\theta_1 = \theta_2$, the total electrical length is minimized:

$$\Theta = \theta_1 + \theta_2 = 2 \cdot \tan^{-1}\left(\sqrt{R}\right).$$

When $R < 1$, $\Theta$ will be less than $\pi/2$. As a result, the physical length of the resonator will be shorter than $\lambda/4$. Another property of SIR is that, its high order harmonics does not appear at $n\omega_0$, but at frequency regime much far away, depending on its impedance ratio [11].

Compared to the UIR, SIR adds a discontinuity at the step edge, which can be modeled as an additional capacitor to the ground [15]. As a result, some non-propagation modes can be launched at the step and finally dissipated by thermal or radiation processes. However, such kind of loss is comparable to that introduced by the open ends of a resonator. From current knowledge about loss mechanism in coplanar resonators [16]–[18], radiation loss is not among the major source of losses (dielectric loss, random interaction with TLS, etc.). It is believed that SIRs have comparable internal quality factor to that of commonly used UIRs.

## II. Experimental Methods

### A. Resonator Design

We designed two patterns, each of which includes several coplanar resonators capacitively coupled to a CPW transmission line. In design I, there are 8 SIRs with the same impedance ratio and 2 UIRs. The geometry of the SIRs was chosen to have the same $(2g + w)$ value for both high and low impedance parts. Here $g$ and $w$ correspond to the gap and trace widths, respectively. The low impedance half has a characteristic impedance of $Z_1 = 50.5\,\Omega$, and the high impedance half has a characteristic impedance of $Z_2 = 92.6\,\Omega$. As a result, the impedance ratio $R = Z_1/Z_2 = 0.54$. The electrical length of both parts were tried to design as the same, so the total length of the resonator will be $\sim$19% shorter than that of uniform impedance design. The purpose of including 2 UIRs in design I was to make an on-chip comparison. Both of them were designed with $Z_0 \sim 50\,\Omega$ with different trace widths.

We expected the internal Q-factors of the SIRs and UIRs to be in the range of $10^5 \sim 10^6$, so we chose the coupling Q-factor ($Q_c$) to be about $2 \times 10^5$, which corresponds to a coupling capacitance to the transmission line of $\sim 0.8\,\text{fF}$ at 6.5 GHz.

In design II, a group of 5 SIRs with $R = 0.41$ was included. Under such a design parameter, the total length of the SIRs can be $\sim$27% shorter than that of UIRs. The area occupied by a SIR in design II is about $330 \times 500\,\mu\text{m}^2$. When compare to the UIRs used in ref [10], an area reduction of 15% can be reached. As a result, SIRs can make the superconducting quantum computing architecture more compact and increase the capability of including more qubits on a single chip.

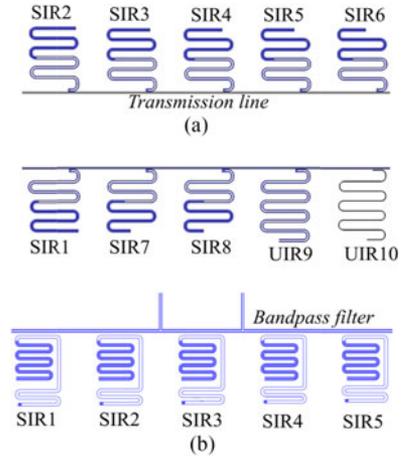

Fig. 2. (a) The layout of resonators in design I. All the resonators are capacitively coupled to a coplanar transmission line with $Z_0 \sim 50\,\Omega$. (b) The layout of resonators in design II. All the resonators are capacitively coupled to a bandpass filter with a bandwidth of $\sim$300 MHz. The detailed the design parameters are listed in Table I.

TABLE I
Design Parameters for the Resonators

| Resonator No. | Geometry ($\mu$m) ($g_1$, $w_1$, $g_2$, $w_2$) | length ($\mu$m) ($l_1$, $l_2$) | Coupling capacitance[a] (in fF) | $Z_1/Z_2$ |
|---|---|---|---|---|
| Design I | | | | |
| SIR1 | 18, 4, 10, 20 | 2151, 2128 | 0.8 | 0.54 |
| SIR2 [b] | 18, 4, 10, 20 | 2114, 2092 | 0.8 | 0.54 |
| SIR3 | 18, 4, 10, 20 | 2083, 2060 | 0.8 | 0.54 |
| SIR4 [b] | 18, 4, 10, 20 | 2048, 2027 | 0.8 | 0.54 |
| SIR5 | 18, 4, 10, 20 | 2018, 1995 | 0.8 | 0.54 |
| SIR6 [b] | 18, 4, 10, 20 | 1984, 1963 | 0.8 | 0.54 |
| SIR7 | 18, 4, 10, 20 | 1956, 1933 | 0.8 | 0.54 |
| SIR8 | 18, 4, 10, 20 | 1929, 1905 | 0.8 | 0.54 |
| UIR9 | 10, 20 | 4702 | – | – |
| UIR10 | 2, 4 | 4569 | – | – |
| Design II | | | | |
| SIR1 | 18, 4, 10, 20 | 1677, 1683 | 1.5 | 0.41 |
| SIR2 | 18, 4, 10, 20 | 1670, 1676 | 1.5 | 0.41 |
| SIR3 | 18,4,10, 20 | 1663, 1669 | 1.5 | 0.41 |
| SIR4 | 18, 4, 10, 20 | 1656, 1662 | 1.5 | 0.41 |
| SIR5 | 18, 4, 10, 20 | 1649, 1655 | 1.5 | 0.41 |
| Design III | | | | |
| UIR1 | 4, 4 | 5156 | – | – |
| UIR2 | 8, 8 | 5136 | – | – |
| UIR3 | 16, 16 | 5156 | – | – |

[a] The coupling capacitance was simulated by Sonnet. As all the SIRs in design I and design II have the same coupling geometry, only one of them was simulated.
[b] Indicates SIRs that have rounded steps.

SIRs in design II were designed for the readout of superconducting qubits. Thus, their coupling Q-factor values are designed to be smaller in the range of 2–10 thousands. In addition, they are coupled to a low-Q resonator ($Q_L \sim 23$) acting as a bandpass filter with a bandwidth of $\sim$300 MHz.

The layouts of both designs are shown in Fig. 2. Detailed parameters for resonators in both designs are listed in Table I.

In practice, all the resonators inevitably have discontinuities including open ends, bends or steps. As discontinuities can generally be modeled as additional capacitance to the ground, they will modify the electrical length because of these finite impedance capacitors [19]. During design, we made corrections to the physical length of the resonators empirically by Sonnet simulations.

### B. Resonator Fabrication

The resonators were fabricated using aluminum thin films deposited on C-plane sapphire substrates ($10 \times 10$ mm$^2$). The thin films were about 100 nm thick and were deposited by an ultra-high vacuum electron-beam evaporation system (base pressure $\sim 10^{-9}$ Torr) at a rate of $\sim 1$ nm/sec. To optimize the metal-vacuum interfaces, resonators, CPW transmission line, band-pass filter and electrical pads were defined by e-beam lithography (100 keV acceleration voltage, 2–10 nA current) in a single step, and then patterned by wet etching.

### C. Resonator Measurement

The resonators were measured by standard transmission response measurement in a Helium-free dilution refrigerator (Bluefors LD250) with a base temperature of $\sim 10$ mK. The sample space was carefully shielded from static magnetic field by one layer of low-temperature $\mu$-metal shield and 3-layers of room-temperature $\mu$-metal shield. A copper can radiation shield was also installed. In addition, non-magnetic SMA connections were used for launch and output ports to avoid local stray field around the samples. The input microwave signal was heavily attenuated by a series of attenuators from room-temperature plate to the input port of the transmission line to suppress the environmental and radiation noise from room-temperature. Output signal was pre-amplified by a HEMT mounted at the 4-K stage, followed by further amplification at room-temperature before being sent into a vector network analyzer (Agilent PNA-N5244A). Two isolators were inserted in series between the tested sample and the HEMT to minimize the back-action from the HEMT. Photographs of low-temperature installation and a schematic of the measurement configuration are shown in Fig. 3.

The internal Q-factor and coupling Q-factor were extracted according to the asymmetry impedance model described in literature [8]. The fitting results show that our measured responses agree quite well with this model. Typical $S21$ response of resonators and fitting results are shown in Fig. 4.

What we are interested in is the internal Q-factor at low excitation power comparable to that generated by a single microwave photon circulating in the resonators, that is, excitation in the quantum region. The average photon number circulating in a resonator can generally be estimated by the following equation [16]:

$$N_{\text{circ}} = P_{\text{inc}} Q_L \frac{(1 - Q_L/Q_i)}{n\pi} / h f_r^2.$$

Where $P_{\text{inc}}$ is the incident power, $Q_L$ is the loaded Q-factor, $n$ is the $n$th harmonic being probed, and $f_r$ is the resonance frequency. For a resonator with a coupling Q-factor of $2 \times 10^5$ and a typical internal Q-factor of $1 \sim 10 \times 10^5$, the incident power for one average photon is approximately $-144$ dBm $\sim -152$ dBm at 6 GHz. We measured the response under different stimulating power from –20 to –65 dBm. While including the total attenuation and losses on the input coaxial lines, the average photon number is from $10^5$ to less than 1 for different resonators.

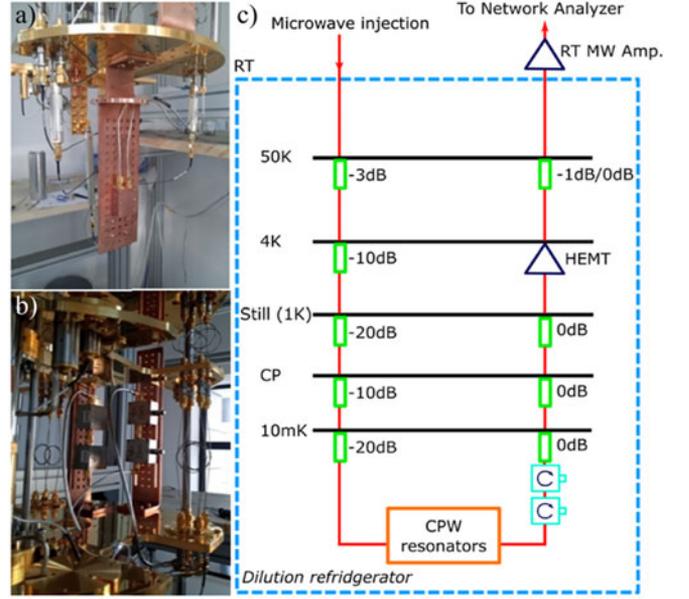

Fig. 3. (a) and (b) The photographs of the mounted sample box and output circuits on the MXC plate. (c) Schematic for the resonator measurement configuration.

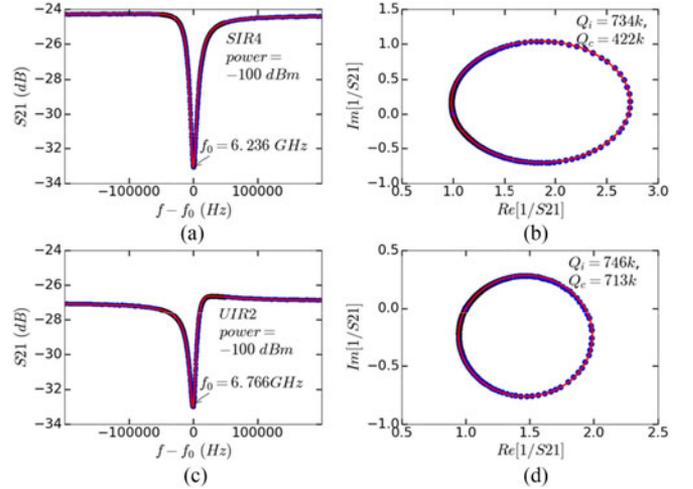

Fig. 4. (a) The measured $S21$ response of SIR4 in device I (blue circle) and the fitting result (red line). The incident power is estimated to $\sim 100$ dB·m. (b) The normalized $\tilde{S}_{21}^{-1}$ shown in a smith chart (blue circle) and the fitting result (red line). (c) and (d) The measurement and fitting results of UIR2 in device I under the same condition as that of SIR4.

## III. RESULTS AND DISCUSSION

Device I was fabricated following the design I. Fig. 5 shows the extracted internal Q-factors of the resonators under different circulating photon numbers. At the high power limit, $Q_i$ values of SIRs and UIRs on the same chip are comparable in the range of 6–10 $\times 10^5$. As the incident power is decreased and approaches to single-photon level, $Q_i$ of two UIRs degrade to about $2 \times 10^5$, while all the SIRs still keep on the level of half-million. For comparison, we also include measurement results from another device (device III, the design parameters are shown in Table I, labeled as design III). Under single-photon level, SIRs also have comparable performance to UIRs in device III.

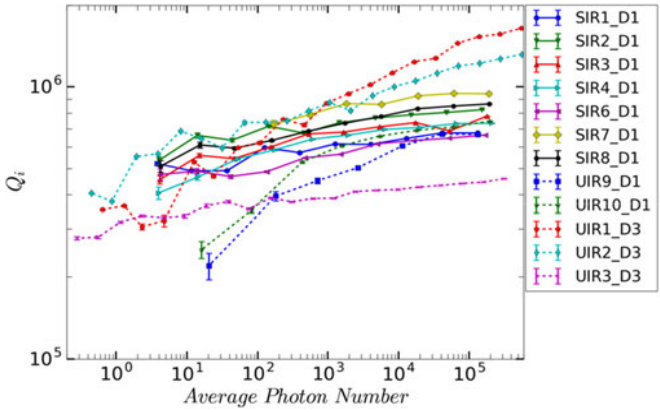

Fig. 5. The power (average photon number) dependence of extracted internal Q-factors of resonators in device I (SIR1-8, UIR9-10_D1) and III (UIR1-3_D3).

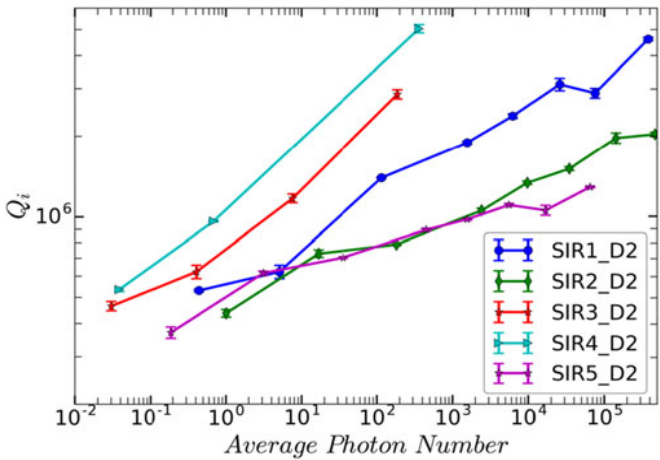

Fig. 6. The power (average photon number) dependence of extracted internal Q-factors of resonators in device II.

Device II was fabricated following the design II. Extracted internal Q-factor values for all the SIRs under different incident powers are shown in Fig. 6. In the high power range, these SIRs exhibit extremely high $Q_i$ values. For SIRs with low $Q_c$ value, it was failed to extract the Q information by using current model. With decreasing the incident power, $Q_i$ decreases fast but is still about half million in the quantum region.

These results indicate that discontinuities introduced by impedance steps in SIRs do not lead to notable degradation of internal Q-factor.

## IV. CONCLUSION

In summary, we have fabricated and measured superconducting microwave coplanar resonators with both the stepped-impedance and uniform-impedance designs. The resonators were made of aluminum films deposited on C-plane sapphire substrates. The measurements were taken at milli-kelvin temperatures in both the many photon and single photon limits. The $Q_i$ values extracted from the measured transmission characteristics (S21) for both types of resonators show that the performance of SIRs is comparable with that of UIRs and is not compromised by the discontinuities presented in SIRs. By choosing proper impedance ratio, the physical length of SIR can be much shorter than that of UIR. These results indicate that SIR can find its applications in quantum computer architectures with more compact design.

ACKNOWLEDGMENT

The authors thank Chengchun Tang for technical assistance with e-beam lithography.